\documentclass[a4paper,10pt]{article}
\usepackage[utf8]{inputenc}
\usepackage{amsmath}
\usepackage{amsfonts}
\usepackage{graphicx}
\usepackage{subcaption}
\usepackage{authblk}

\newcommand{\beginsupplement}{%
        \setcounter{table}{0}
        \renewcommand{\thetable}{S\arabic{table}}%
        \setcounter{figure}{0}
        \renewcommand{\thefigure}{S\arabic{figure}}%
     }

\title{Diagnosing model misspecification and performing generalized Bayes' updates via probabilistic classifiers}
\author[1]{Owen Thomas}
\author[2]{Jukka Corander}
\affil[1,2]{Department of Biostatistics, University of Oslo}
\begin{document}

\maketitle

\begin{abstract}
Model misspecification is a long-standing enigma of the Bayesian inference framework as posteriors tend to get overly concentrated on ill-informed parameter values towards the large sample limit. Tempering of the likelihood has been established as a safer way to do updates from prior to posterior in the presence of model misspecification. At one extreme tempering can ignore the data altogether and at the other extreme it provides the standard Bayes' update when no misspecification is assumed to be present. However, it is an open issue how to best recognize misspecification and choose a suitable level of tempering without access to the true generating model. Here we show how probabilistic classifiers can be employed to resolve this issue. By training a probabilistic classifier to discriminate between simulated and observed data provides an estimate of the ratio between the model likelihood and the likelihood of the data under the unobserved true generative process, within the discriminatory abilities of the classifier. The expectation of the logarithm of a ratio with respect to the data generating process gives an estimation of the negative Kullback-Leibler divergence between the statistical generative model and the true generative distribution. Using a set of canonical examples we show that this divergence provides a useful misspecification diagnostic, a model comparison tool, and a method to inform a generalised Bayesian update in the presence of misspecification for likelihood-based models.
\end{abstract}

\section{Introduction}

Bayesian inference is one of the cornerstones of modern model-based statistics, having experienced a renaissance after the popularization of Monte Carlo methods since early 1990's. It is fundamentally constrained to operate over a family of models contained within the prior distribution asserted by an expert \cite{bernardo2009bayesian}. From the pragmatic perspective, we can almost never assume that the hypothetical true data generative process falls within a specified pool of prior models, which implies that model misspecification remains a frequent concern. A classical example of Bayesian misspecification is the asymptotic posterior convergence to the single parameter value minimizing the Kullback-Leibler (KL) divergence to the true model, even when such a model is a poor characterisation of reality \cite{bernardo2009bayesian}.

There exists a large literature concerned with misspecification under the Bayesian inference framework, much of it being based on the use of tempered likelihoods \cite{bissiri2016general,grunwald2017inconsistency,dalalyan2008aggregation}. When a model misspecification is suspected, then a Bayes' update is performed with the likelihood raised to a power $0<t<1$, termed as tempering. This is done to avoid overly confident convergence to a poor model, leading to a generalised posterior distribution. Despite recent progress on such safer Bayesian inference procedures \cite{grunwald2017inconsistency}, it is generally an open problem how to best diagnose model misspecification and to  decide on whether tempering is useful enough and which value of $t$ should preferably be used. We aim at answering these questions by employing probabilistic classifiers.

As shown in the literature, there exists a fundamental link between probabilistic classification and density ratio estimation \cite{sugiyama2012density}. When we are interested in estimating a ratio of two distributions of interest, it is possible to do this indirectly by sampling from each distribution and then training a probabilistic classifier to discriminate between the two populations of data. The odds ratio of the trained classifier evaluated at any new samples provides a principled approximation of the ratio of their corresponding distributions. Such methods have been used extensively in the context of machine learning and sampling-based inference \cite{sugiyama2012density}.

Ratio estimation with the aid of classifiers in various incarnations has received considerable attention in the context of likelihood-free inference. For example, classifier-based ABC trained a discriminator between simulated and observed data to provide a discrepancy measure for ABC rejection \cite{gutmann2018likelihood}. As example of frequentist likelihood-free inference, the CARL method evaluates two likelihoods, each conditional on a given parameter value, by calibrated classifiers \cite{cranmer2015approximating}. Already earlier Pratola et al. used similarly non-parametric estimates of likelihood ratios to calibrate computer simulation model parameters for complex spatio-temporal models \cite{pratola2013fast}. The recent LFIRE method evaluates a likelihood to evidence ratio to perform a pointwise Bayesian update \cite{thomas2016likelihood,kokko2019pylfire}, and more recent work has used a global classifier to approximate a likelihood to evidence ratio to perform a global Bayesian update \cite{hermans2019likelihood}.

Principled use of classifiers for implicit generative model fitting is also widespread within machine learning, the most famous version being Generative Adversarial Networks (GANs), which use a discriminative neural network as a loss function for a generative neural network by classifying between the generated simulated data and the observed data \cite{goodfellow2014generative}. Such applications of classifiers has proved useful in improving the inference for such models, and their probabilistic foundations and relevance to likelihood-free inference have been explored by later work \cite{mohamed2016learning,tran2017hierarchical}.

\section{Bayesian inference under misspecifcation}

A central approach to handling misspecified models is to use tempered likelihoods. When a model misspecification is suspected, then an update from prior to posterior is performed with a likelihood raised to a power $0<t<1$, to avoid overly confident convergence to a poor model. This produces a generalised posterior distribution $p_t(\theta|X)$ defined as

\begin{equation}
 p_t(\theta|X) = \frac{p(X|\theta)^{t} p(\theta)}{p_t(X)},
\end{equation}

where $X$ are the observed data, $\theta$ are the model parameters and $p_t(X)$ is the generalized evidence, or the marginal likelihood under the tempered model. For the particular choice $t=1$ a full Bayesian update is recovered and correspondingly for $t=0$ the prior is returned. The values of $t$ in between represent fractional updates between the two extremes, while still using the entire data set $X$. This can be interpreted as ensuring that the epistemological uncertainty from a vague belief matches the uncertainty exhibited by the data, even when the individual models are not good reflections of reality. Several methods exist to choose a tempering level, including the SafeBayesian \cite{grunwald2017inconsistency}, Generalised Bayes Updates \cite{bissiri2016general} and PAC-Bayesian reasoning \cite{dalalyan2008aggregation}.

SafeBayesian inference considers the generalised posterior distributions with likelihood tempered with a parameter $t$ \cite{grunwald2017inconsistency}. The optimal level of tempering is determined by minimising an expected log loss between the observed data and data simulated from the generalised posterior for each value of $t$. Such a method increases the robustness of the inference against model misspecification by allowing a model to perform a smaller update if the posterior is at risk of over-concentration.

Misspecification has also received specific attention within Approximate Bayesian Computation when considering choice of summary statistics and rejection thresholds \cite{frazier2017model}. Similarly, a theoretical synthesis of PAC-Bayesian theory and likelihood-free inference has been achieved \cite{ridgway2017probably}. It has been suggested that performing inferences with discrepancies and data summaries rather than explicit likelihoods may reduce the risk of misspecification and increase the robustness of the inference.

\section{CARMEN method for model evaluation}

Training a classifier to recognize the difference between the data predicted by a model (i.e. using generated data) and the observed data provides an estimate of the ratio of the predictive distribution of the model $\mathcal{M}$ and the generative process under the true model $\mathcal{T}$, conditional on the space of models that can be discriminated by the classifier used $\mathcal{C}$. More specifically the ratio is estimated by the predictive odds ratio of a probabilistic classifier $\mathcal{C}$ trained using the observed data $X_{v}$, implicitly drawn from a true generative process $p_{\mathcal{T}}(X)$, and simulated data $X_{s}$ drawn from the predictive distribution of a statistical model $p_{\mathcal{M}}(X)$. We denote this ratio by

\begin{equation}
Z_{X,\mathcal{M},\mathcal{C}} = \frac{p_{\mathcal{M}| \mathcal{C}}(X)}{p_{\mathcal{T}|\mathcal{C}}(X)},
\end{equation}

with an explicit conditioning on the classifier $\mathcal{C}$. When evaluated on an observed data set $X_{v}$, the expectation of the logarithm of the ratio converges to the negative KL divergence between the statistical model and the true model for increasing amounts of observed data, conditional on the classifier. Assuming the $n$ observed sampler are i.i.d. we thus have 

\begin{align}
 \log Z_{X_{v},\mathcal{M},\mathcal{C}} =& \frac{1}{n}\sum_{i=1}^{n} \log \frac{p_{\mathcal{M}|\mathcal{C}}(X_{v}^{i})}{p_{\mathcal{T}|\mathcal{C}}(X_{v}^{i})}\\
  \lim_{n\to\infty} \frac{1}{n}\sum_{i=1}^{n} \log \frac{p_{\mathcal{M}|\mathcal{C}}(X_{v}^{i})}{p_{\mathcal{T}|\mathcal{C}}(X_{v}^{i})}    = &\mathbb{E}_{p_{\mathcal{T}}(X)} \left[\log\frac{p_{\mathcal{M}|\mathcal{C}}(X)}{p_{\mathcal{T}|\mathcal{C}}(X)}\right]\nonumber\\
  =&\int p_{\mathcal{T}}(X)\left(\log\frac{p_{\mathcal{M}|\mathcal{C}}(X)}{p_{\mathcal{T}|\mathcal{C}}(X)}\right)dX\nonumber\\
  = &- D_{KL}(p_{\mathcal{T}|\mathcal{C}} || p_{\mathcal{M}|\mathcal{C}})
\end{align}\label{logratio}

We have assumed the true generative distribution $p_{\mathcal{T}}$ and the classifier conditional true distribution $p_{\mathcal{T}|\mathcal{C}}$ to be equivalent in the definition of the KL divergence. Evaluating the expected log ratio of the predictive distribution of the model under consideration with the likelihood under the generative process is a useful and interpretable diagnostic for model misspecification assessment and model comparison, and equivalent to estimating the KL divergence between the true generative model and the statistical model. It is worth noting that if the log odds ratio is instead evaluated on simulated data, this is equivalent to an estimate of the reverse KL divergence, i.e.  $D_{KL}(p_{\mathcal{M}|\mathcal{C}} || p_{\mathcal{T}|\mathcal{C}})$. This is less widely used in the analysis of misspecification than the version considered in the remainder of the article.

We also note that no further assumptions are made about the nature of the true generative process, other than that the classifier is able to discriminate between the simulated data and the observed data. As is customary in classifier training, it is also recommended here to use cross-validation to stabilise the estimation of the expected log ratios, which increases the robustness of the estimate and also removing the need to strictly partition observed data into a training and a test set. We call the above method CARMEN: Classification to Assess Ratios for Misspecification Estimation and Negotiation.

\section{Testing for misspecification}\label{MisspecTest}

The ratio introduced in the previous section provides an estimate of model misspecification in more absolute terms than previously possible, by effectively considering the space of all models discriminable by the classifier instead of those considered within the prior or discriminable by the likelihood function. For a sufficiently expressive classifier, we thus expect to arrive at a useful method for misspecification assessment. A log ratio of approximately zero corresponds to a KL divergence of approximately zero between the statistical simulator and the true model, which implies that the model is well-specified. In contrast, very large negative values of the log ratio indicate that our model is not an adequate representation of the data. As far as we know, such an analysis would not be directly possible within the framework of ordinary likelihood-driven Bayesian inference.

The summation operation in the expected log ratio in Equation \ref{logratio} also provides a method of assessing the uncertainty associated with the estimate of the KL. With a large amount of validation data $X_{v}$, and assuming that the individual log-ratio approximations on each data point $Z_{X_{v}^{i}}$, $\log Z_{X_{v}^{i},\mathcal{M},\mathcal{C}}$, are independently and identically distributed with finite mean and variance, we expect the distribution of their sum to converge to a univariate Gaussian distribution by the central limit theorem:

\begin{align}
\log Z_{X_{v},\mathcal{M},\mathcal{C}}=\sum_{i=1}^{n} \log Z_{X_{v}^{i},\mathcal{M},\mathcal{C}} \xrightarrow{d} \mathcal{N}(\mu_{\log Z},\sigma_{\log Z}^{2})
\end{align}

The distribution $\mathcal{N}(\mu_{\log Z},\sigma_{\log Z}^{2})$ allows us to quantify the support against the divergence equalling zero when dealing with noisy data or a suboptimally calibrated classifier. The parameters $\mu_{\log Z}$ and $\sigma_{\log Z}^{2}$ can be determined by estimating the distribution of the individual log ratio estimates $\log Z_{X_{v}^{i},\mathcal{M},\mathcal{C}}$.

It is possible to use a one-tailed likelihood-ratio test for the hypothesis that the KL divergence is zero, as we expect the summation operation to make a one-tailed t-test relatively robust for large values of $n$ by the central limit theorem. Formally:

\begin{align}
    H_{0}:\quad \log Z_{X_{v},\mathcal{M},\mathcal{C}} = 0\\
    H_{1}:\quad \log Z_{X_{v},\mathcal{M},\mathcal{C}} < 0
\end{align}

The one-tailed property follows from the fact that values of $\log Z_{X_{v},\mathcal{M},\mathcal{C}}>0$ have no proper interpretation beyond noise, but would reasonably be interpreted as a noisy evaluation of an estimated KL divergence equal to zero. The t-statistic is calculated from the individual log ratio estimates $\log Z_{X_{v}^{i},\mathcal{M},\mathcal{C}}$, using their sample mean $\bar{x}$ and standard deviation $s$:

\begin{align}
t_{\log Z = 0}=\frac{\bar{x}}{s/\sqrt{n}}
\end{align}

The one-tailed test is then performed with $n-1$ degrees of freedom. A small p-value can be interpreted to indicate a poorly specified model: there is strong evidence for the alternative hypothesis that the model is misspecified. However, a large p-value should still be interpreted with some caution, as there may not be enough validation data or the space of models discriminable by the classifier may not be large enough. In a situation where the central limit theorem is not expected to be valid, a nonparametric test such as the Wilcoxon signed-rank test may be considered more appropriate. 

\section{CARMEN for generalised belief updates}

With the aid of data partitioning, it is possible to use CARMEN to assess an appropriate likelihood tempering level for a generalised Bayesian belief update. Separate partitions of observed data are then used for the Bayesian update $X_u$ and the validation step with classification $X_v$.

Our approach to generalized belief updates builds on that of the SafeBayesian \cite{grunwald2017inconsistency}, which considers predictive data from each tempering level of a generalised Bayesian update using a loss evaluated against the observed data. In this work, an extra step considers the classifier-approximated values of the expected log ratio $\log Z_{X_{v},\mathcal{M},\mathcal{C}}$ for different values of tempering $t$, hereafter notated $\log Z_{X_{v},\mathcal{M},\mathcal{C},t}$, allowing for an analysis of the model specification as a function of the tempering.

It is possible to identify the most appropriate level of tempering by maximising the predictive log probability of the data under different levels of tempering, rather than evaluating $\log Z_{X_{o},\mathcal{M},\mathcal{C},t}$ for many levels of tempering. The use of the predictive log probability is fundamentally related to the SafeBayesian approach, in which an expected loss is evaluated for each value of $t$. The optimisation of the predictive log probability to find $t$ requires less computation, demanding a one-dimensional optimisation of the tempering quantity $t$, followed by a single classifier-based estimation of the log-ratio. In principle, the log ratio only varies as a function of $t$ in the numerator, so the two estimates should be maximised by the same value of $t$.

The predictive distribution of the fractional posterior $p_t(\theta|X_{u})$ on a validation data set $X_{v}$ can be specified as a function of the tempered and non-tempered likelihoods according to

\begin{equation}
 p_t(X_{v}|X_{u}) = \int p(X_{v}|\theta) p_t(\theta|X_{u}) d \theta = \int \frac{p(X_{v}|\theta) p(X_{u}|\theta)^{t} p(\theta)d \theta }{p_t(X_{u})}
\end{equation}

Such a tempered predictive distribution is generally analytically accessible for exponential family models.  Having maximised the tempered posterior predictive distribution, it is possible to use the CARMEN classifier to provide an estimate of the KL divergence between the true model and the distribution of models described by the partial Bayesian update, and also a mechanism for testing whether the KL divergence is positive. This is a useful check for misspecification, providing information external to the likelihood function to assess the success of the model fit, and for an interpretable classifier potentially suggesting which data features are causing the misspecification.

\section{Examples of CARMEN}

Here we provide examples of misspecification diagnosis and generalised belief updates for commonly considered canonical models with conjugate priors.

\subsection{Misspecified univariate Gaussian}

\begin{figure}
 \centering
     \begin{subfigure}[t]{0.45\textwidth}
        \includegraphics[width=\textwidth]{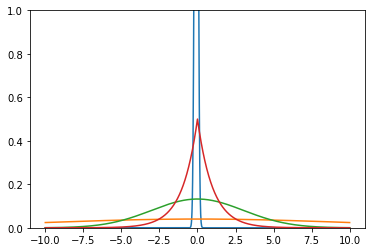}
        \caption{Predictive cdfs.}
        \label{fig:postsgauss}
    \end{subfigure}
    ~ %add desired spacing between images, e. g. ~, \quad, \qquad, \hfill etc. 
     \begin{subfigure}[t]{0.45\textwidth}
        \includegraphics[width=\textwidth]{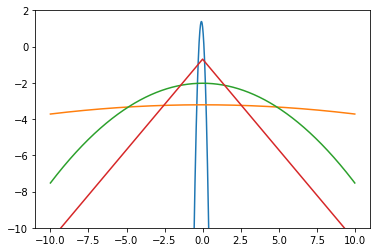}
        \caption{Predictive log cdfs.}
        \label{fig:logpostsgauss}
    \end{subfigure}
    \caption{Predictive distributions associated with the univariate Gaussian example. Orange - overdispersed mean, Blue - underdispersed posterior, Green - CDF of a Gaussian with $\sigma = 3$, Red - CDF of a Laplace distribution with $b=2.13$}\label{fig:gaussdistributions}
\end{figure}

The first example considers a Gaussian model for data with a small variance and a heavily dispersed Gaussian prior over the mean. The resulting prior predictive is very dispersed, quickly converging to a heavily concentrated posterior predictive distribution. The predictive distributions used for all the Gaussian examples are shown in Figure \ref{fig:gaussdistributions}.

\begin{align}
p(X|\mu)=\mathcal{N}(\mu,0.1^2)\\
p(\mu)=\mathcal{N}(0,9.9^2)
\end{align}

\subsubsection{Gaussian with Gaussian true model}\label{sec:gaussgauss}

In this example, the above Gaussian statistical model is combined with a true generative model of a Gaussian distribution of mean zero and standard deviation between that of the prior predictive and the posterior predictive distributions: 

\begin{align}
X_{o}\sim\mathcal{N}(0,3.01^2)
\end{align}

Crucially, the generalised Bayesian posterior predictive distribution is here able to pass through the correct model for a tempering value approximately equal to $10^{-6}$. In our examples 1000 data points are generally used for an update and additional 1000 for validation. Standard logistic regression is used as the classifier, with the summary statistics $X$ and $X^2$ used as the covariates. 10-fold cross validation is used to provide estimates of $\log Z_{X_{v},\mathcal{M},\mathcal{C},t}$ for every member of validation data set. In this example with a known true model, the true value of the log ratio $\log Z_{X_{v},\mathcal{M},\mathcal{C},t}$ is known for all values of $t$. However, the classifier-based approximation is also calculated for all values of $t$ for illustrative purpose.

\begin{figure}[h]
    \centering
    \begin{subfigure}[t]{0.45\textwidth}
        \includegraphics[width=\textwidth]{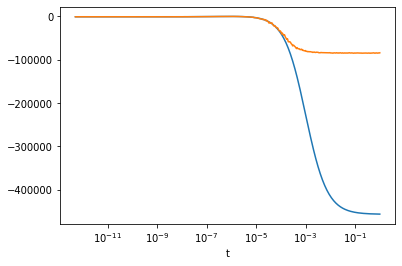}
        \caption{$\log Z_{X_{v},\mathcal{M},\mathcal{C},t}$ for all $t$ up to a full Bayes update.}
    \end{subfigure}
    ~ %add desired spacing between images, e. g. ~, \quad, \qquad, \hfill etc. 
    \begin{subfigure}[t]{0.45\textwidth}
        \includegraphics[width=\textwidth]{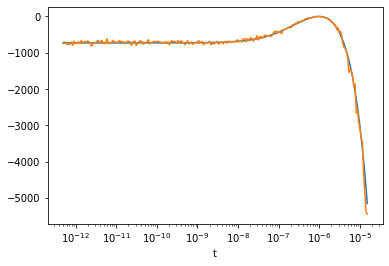}
        \caption{$\log Z_{X_{v},\mathcal{M},\mathcal{C},t}$ for $t$ range around the optimum.}
    \end{subfigure}
\newline
        \begin{subfigure}[t]{0.45\textwidth}
        \includegraphics[width=\textwidth]{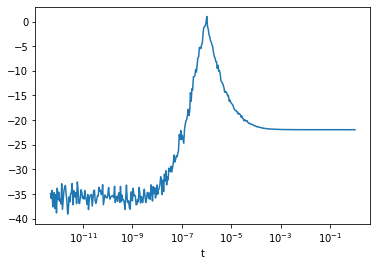}
        \caption{t-statistics.}
    \end{subfigure}
    ~ %add desired spacing between images, e. g. ~, \quad, \qquad, \hfill etc. 
    \begin{subfigure}[t]{0.45\textwidth}
        \includegraphics[width=\textwidth]{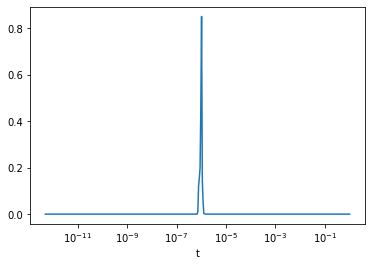}
        \caption{p-values}
    \end{subfigure}
 \caption{The true log ratio $\log Z_{X_{v},\mathcal{M},\mathcal{C},t}$ in orange and classifier-driven approximate in blue for a Gaussian statistical model with Gaussian true distribution, above the t-statistics and associated p-values analysing whether the approximate log ratio is greater than zero.}
\label{fig:gaussgauss}
\end{figure}

Figure \ref{fig:gaussgauss} shows the true and approximated log ratios for different values of tempering $t$. The tempered log predictive probability of the observed data is maximised at $t=9.5 \cdot 10^{-7}$. We observe that the approximation is highly accurate around the maximum of the log ratio, with a bias becoming evident for very large negative values of the log ratios. The large negative values of the log ratio would not be identified as optimal levels of $t$ and as such the bias does not affect the analysis at the optimal level of tempering. The bias represents the inability of the relatively simple classifier to generate exactly calibrated odds ratios in instances of extreme misspecification. Notably, with the optimally tempered generalised posterior predictive at $10^{-6}$, the classifier is well specified and the resulting odds ratio is well calibrated.

Figure \ref{fig:gaussgauss} also shows the test statistics and corresponding p-value from a t-test with a null hypothesis that the sum log ratio is equal to zero. We see that the t-statistic and p-values clearly take a maximum near $10^{-6}$, corresponding to the level of tempering maximising the log predictive probability of $X_v$, we find a p-value of 0.456. Within a frequentist framework, this implies that the null hypothesis of the model being misspecified is not rejected for this value of $t$. However, we do not necessarily advocate making decisions directly from the t-statistics or corresponding p-values as a function of the tempering, as this would introduce a multiple-testing problem. The values are mainly provided here to illustrate the statistical behavior of the approach.

\subsubsection{Gaussian with Laplace Truth}

The second example uses the same statistical model as Section \ref{sec:gaussgauss}, but with a different ground truth corresponding to a Laplace distribution with mean of zero and scale of 2.13, also presented in Figure \ref{fig:gaussdistributions}.

\begin{align}
X_{o}\sim Laplace(b=2.13)
\end{align}

The variance of the generated data under the true process is again equal to three, so the generalised tempered posterior predictive would be expected to most closely approach the true generative process at a tempering value of $10^{-6}$. In contrast to the previous example, the tempered predictive will never exactly overlap with the true generative process owing to the heavier tails of the Laplace distribution. Consequently, model tempering will be expected to provide an improvement relative to a full Bayesian update, but not enable convergence to the true model.

Logistic regression with 10-fold cross validation was similarly used as the classifier, with $X$, $X^2$ and $\ln|X|$ used as the covariates. The logarithm of the data was used to enable the linear classifier to discriminate between the tail behaviour of the data. The same distribution of tempering values was used as previously.

\begin{figure}
    \centering
    \begin{subfigure}[t]{0.45\textwidth}
        \includegraphics[width=\textwidth]{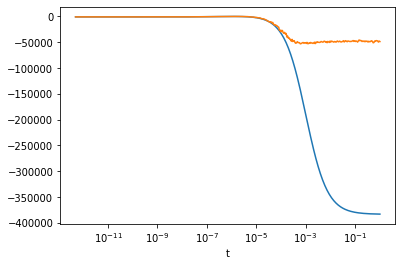}
        \caption{$\log Z_{X_{v},\mathcal{M},\mathcal{C},t}$ for all $t$ up to a full Bayes update.}
    \end{subfigure}
    ~ %add desired spacing between images, e. g. ~, \quad, \qquad, \hfill etc. 
    \begin{subfigure}[t]{0.45\textwidth}
        \includegraphics[width=\textwidth]{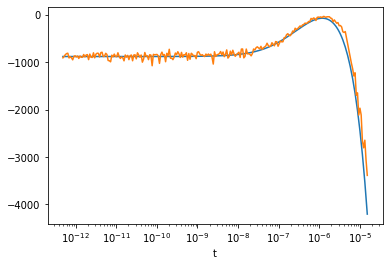}
        \caption{$\log Z_{X_{v},\mathcal{M},\mathcal{C},t}$ for $t$ range around the optimum.}
    \end{subfigure}
    \newline
        \begin{subfigure}[t]{0.45\textwidth}
        \includegraphics[width=\textwidth]{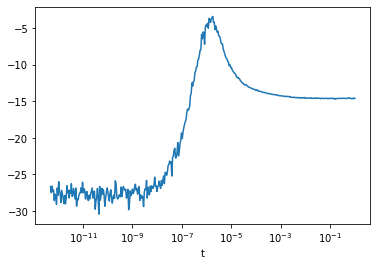}
        \caption{t-statistics.}
    \end{subfigure}
    ~ %add desired spacing between images, e. g. ~, \quad, \qquad, \hfill etc. 
    \begin{subfigure}[t]{0.45\textwidth}
        \includegraphics[width=\textwidth]{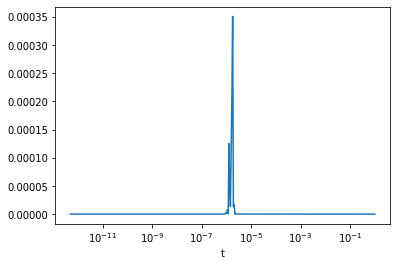}
        \caption{p-values}
    \end{subfigure}
  \caption{The true log ratio $\log Z_{X_{v},\mathcal{M},\mathcal{C},t}$ in orange and classifier-driven approximate in blue for a Gaussian statistical model with Laplace true distribution, above the t-statistics and associated p-values analysing whether the approximate log ratio is greater than zero.}
\label{fig:gausslaplace}
\end{figure}

We see from Figure \ref{fig:gausslaplace} that the classifier-based ratio approximation generates a reasonable approximation to the true log-likelihood near the maximum, capturing the location of the optimal tempering value, while it exhibits a bias for large negative values of the log ratio.  For this example, neither the true log likelihood ratio nor the approximation approach zero, as none of the tempered posterior predictive distributions approach the true generative model. The maximum value of the classifier-driven approximate ratio was -34.9, while the maximum value of the true log likelihood ratio was -71.8.

In Figure \ref{fig:gausslaplace} the t-statistics and corresponding p-values are plotted for a test of whether the expected log ratio is equal to zero. None of the t-statistics approach zero and consequently the corresponding p-values remain far below any conventional threshold of significance. The value of tempering $1.17 \cdot 10^{-6}$ determined by maximising the tempered posterior predictive probability is assigned a p-value of $3.24 \cdot 10^{-6}$ and hence the method has correctly identified that the model is very likely misspecified.

\subsection{Misspecified Poisson Distribution}

We now extend the analysis to discrete data model and start with a Poisson distribution, with a Gamma distribution assumed over the rate parameter, which corresponds to a negative-binomial predictive distribution. The predictive distribution converges to a Poisson distribution in the limit of a large amount of data, but the heavier dispersion of the negative-binomial may make a tempered update preferable.

\begin{align}
p(X|\lambda)=&Poisson(\lambda)\\
 p(\lambda)=&\Gamma(\alpha=3,\beta=0.05)
\end{align}

\begin{figure}
 \centering
     \begin{subfigure}[t]{0.45\textwidth}
        \includegraphics[width=\textwidth]{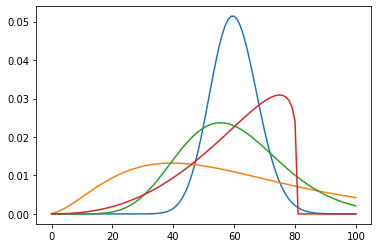}
        \caption{Predictive cdfs.}
        \label{fig:postspoisson}
    \end{subfigure}
    ~ %add desired spacing between images, e. g. ~, \quad, \qquad, \hfill etc. 
     \begin{subfigure}[t]{0.45\textwidth}
        \includegraphics[width=\textwidth]{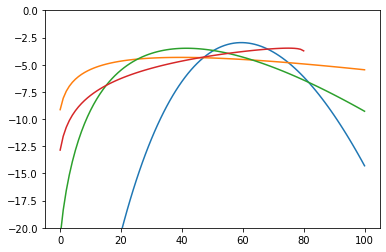}
        \caption{Predictive log cdfs.}
        \label{fig:logpostspoisson}
    \end{subfigure}
    \caption{Smoothed predictive distributions associated with the Poisson example. Orange - the overdispersed mean, Blue - the underdispersed posterior, Green - the negative binomial model used in Section \ref{sec:PoissonNegativeBinomial}, Red - CDF of a Beta binomial distribution used in Section \ref{sec:PoissonBetaBinomial}}\label{fig:poissonposts}
\end{figure}

 The corresponding prior and posterior distributions associated with this example are presented in Figure \ref{fig:poissonposts}, with the true generative distributions used for the examples below.

\subsubsection{Poisson with Negative Binomial Truth}\label{sec:PoissonNegativeBinomial}

In this section we consider observed data drawn from a negative binomial model according to:

\begin{align}
X_{o}\sim NB(r=63,p=0.488)
\end{align}

In this example, the generalised posterior predictive distribution is expected to pass through the exact generative distribution for a tempering of $10^{-3}$.

\begin{figure}
    \centering
    \begin{subfigure}[t]{0.45\textwidth}
        \includegraphics[width=\textwidth]{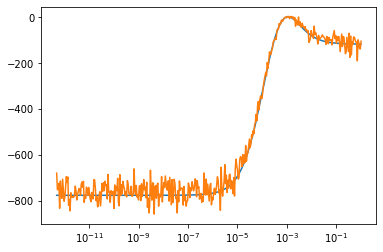}
        \caption{$\log Z_{X_{v},\mathcal{M},\mathcal{C},t}$ for all $t$ up to a full Bayes update.}
    \end{subfigure}
    ~ %add desired spacing between images, e. g. ~, \quad, \qquad, \hfill etc. 
    \begin{subfigure}[t]{0.45\textwidth}
        \includegraphics[width=\textwidth]{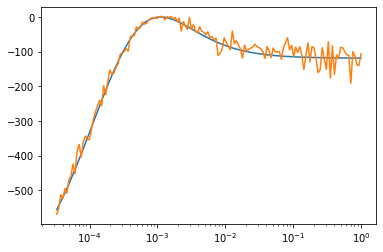}
        \caption{$\log Z_{X_{v},\mathcal{M},\mathcal{C},t}$ for $t$ range around the optimum.}
    \end{subfigure}
    \newline
        \begin{subfigure}[t]{0.45\textwidth}
        \includegraphics[width=\textwidth]{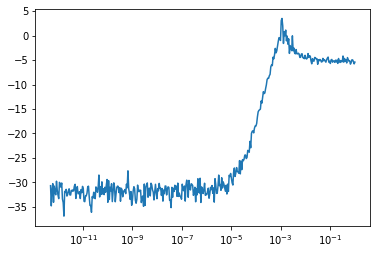}
        \caption{t-statistics.}
    \end{subfigure}
    ~ %add desired spacing between images, e. g. ~, \quad, \qquad, \hfill etc. 
    \begin{subfigure}[t]{0.45\textwidth}
        \includegraphics[width=\textwidth]{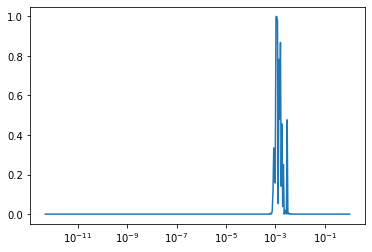}
        \caption{p-values}
    \end{subfigure}
    \caption{The true log ratio $\log Z_{X_{v},\mathcal{M},\mathcal{C},t}$ in orange and classifier-driven approximate in blue for a Poisson statistical model with negative-binomial true model, above the t-statistics and corresponding p-values of a one-tailed t-test evaluating whether the sum log ratio is greater than zero}\label{fig:poissonNB}
\end{figure}

Logistic regression classifier was trained using the following summary statistics: $X$, $X^2$, $X^3$ and $X^4$. We can see in Figure \ref{fig:poissonNB} that the classifier estimation of $\log Z_{X_{v},\mathcal{M},\mathcal{C},t}$ accurately captures the true log ratio, with both approaching zero at their maximum with a tempering value of $1.1 \cdot 10^{-3}$. The maximum values of the true and approximate log ratios were 0.778 and 2.30, respectively.

Figure \ref{fig:poissonNB} displays the t-statistics and corresponding p-values determining whether the expected log ratio is equal to zero. The  p-value associated with tempering from the optimised generalised predictive posterior is 0.9998, indicating negligible evidence that the model is misspecified.

\subsubsection{Poisson with Beta-Binomial Truth}\label{sec:PoissonBetaBinomial}

Here we consider the same Poisson model with a Beta-binomial true generative distribution, with a support defined up to 80 trials. As such, it may be underdispersed relative to the assumed statistical model defined over all positive integers:

\begin{align}
X_{obs}\sim Beta\mathcal{B}(a=41.75,b=78.25,n_{bb}=80)
\end{align}

\begin{figure}
    \centering
    \begin{subfigure}[t]{0.45\textwidth}
        \includegraphics[width=\textwidth]{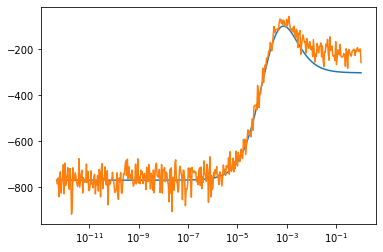}
        \caption{$\log Z_{X_{v},\mathcal{M},\mathcal{C},t}$ for all $t$ up to a full Bayes update.}
    \end{subfigure}
    ~ %add desired spacing between images, e. g. ~, \quad, \qquad, \hfill etc. 
    \begin{subfigure}[t]{0.45\textwidth}
        \includegraphics[width=\textwidth]{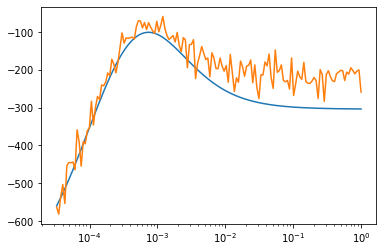}
        \caption{$\log Z_{X_{v},\mathcal{M},\mathcal{C},t}$ for $t$ range around the optimum.}
    \end{subfigure}
    \newline
        \begin{subfigure}[t]{0.45\textwidth}
        \includegraphics[width=\textwidth]{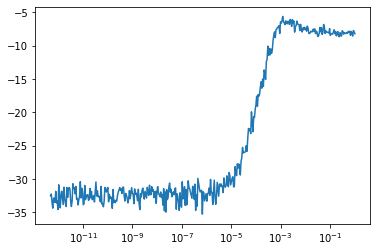}
        \caption{t-statistics.}
    \end{subfigure}
    ~ %add desired spacing between images, e. g. ~, \quad, \qquad, \hfill etc. 
    \begin{subfigure}[t]{0.45\textwidth}
        \includegraphics[width=\textwidth]{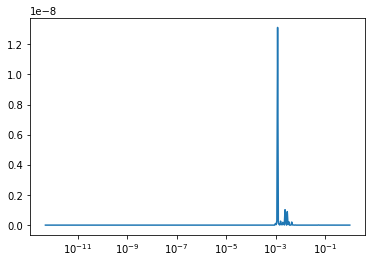}
        \caption{p-values}
    \end{subfigure}
    \caption{The true log ratio $\log Z_{X_{v},\mathcal{M},\mathcal{C},t}$ in orange and classifier-driven approximate in blue for a Poisson statistical model with Beta-binomial true model, above the t-statistics and corresponding p-values of a one-tailed t-test evaluating whether the sum log ratio is greater than zero.}\label{fig:poissonBetaB}
\end{figure}

We used the same classifier and summary statistics as Section \ref{sec:PoissonNegativeBinomial}. Figure \ref{fig:poissonBetaB} illustrates that CARMEN provides a reasonable approximation to the true log ratio $\log Z_{X_{v},\mathcal{M},\mathcal{C},t}$ across all values of the tempering, with the maximum approximate log ratio occurring at $t=1.2 \cdot 10^{-3}$, and the log predictive posterior maximised for $t=7.4 \cdot 10^{-4}$. The maximum values themselves are -58.65 and -100.62 respectively, suggesting a poorly specified model. This conclusion is confirmed by the t-statistics in Figure \ref{fig:poissonBetaB}, all of which indicate strong evidence for misspecification. The values of tempering derived from maximising the expected log generalised predictive posterior is assigned a p-value of $3.5 \cdot 10^{-13}$, demonstrating that even the optimally tempered model is still very likely misspecified.

\subsection{Misspecified Linear Regression}\label{sec:regression}

Linear regression is widely used as an exploratory option for supervised learning, but in practice it generally leads to misspecified models for real data sets. Here we provide a framework for performing a tempered Bayesian update for linear regression and assessing the misspecification level of the resulting generalised Bayesian posterior predictive.

We consider a univariate regression model, although the framework generalises in a fairly straightforward manner to multivariate regression. Normal Inverse Gamma priors were placed on the regression coefficients $\theta$ and on the regression noise variance $\sigma^2$:

\begin{align}
p(y|X,\theta,\sigma^2)=&\mathcal{N}(\theta^T X,\sigma^2)\\
p(\theta,\sigma^{2})=&\mathcal{N}\Gamma^{-1}(\theta_0=0,n_0=1,\alpha=2.,\beta=2.)
\end{align}

The priors used result in a student-t generalised posterior predictive distribution, converging to a Gaussian predictive in the limit of a large amount of data. As such, a tempered Bayesian update may be appropriate in the case of misspecification to better capture the dispersion present in the data. 

\subsubsection{Linear Regression with t-noise truth}

Here we consider the situation in which the true data is generated from a linear model with t-distributed noise.

\begin{align}
y_{o}\sim \mathcal{T}(loc=X_{o},scale=1.22,df=3)
\end{align}

In this situation, the true model may fall within the path of a partial Bayesian update, resulting in a well-specified model with a tempered Bayesian update. Because of the multivariate nature of the posterior predictive distribution, it is not guaranteed that the tempered inference pathway will pass through the correct model, but we might expect a partial update to capture the dispersion in the data set more accurately than a fully converged Gaussian posterior predictive. Logistic regression classifier was used with summary statistics: $|y|$, $y^2$, $\ln|y|$ and $yX$.

\begin{figure}
    \centering
    \begin{subfigure}[t]{0.45\textwidth}
        \includegraphics[width=\textwidth]{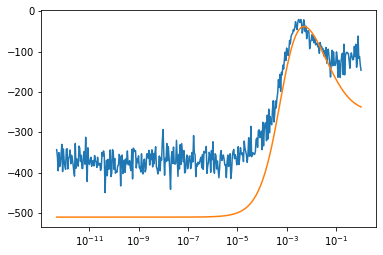}
        \caption{$\log Z_{X_{v},\mathcal{M},\mathcal{C},t}$ for all $t$ up to a full Bayes update.}
    \end{subfigure}
    ~ %add desired spacing between images, e. g. ~, \quad, \qquad, \hfill etc. 
    \begin{subfigure}[t]{0.45\textwidth}
        \includegraphics[width=\textwidth]{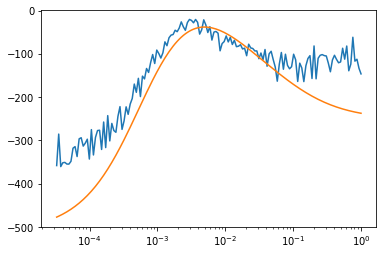}
        \caption{$\log Z_{X_{v},\mathcal{M},\mathcal{C},t}$ for $t$ range around the optimum.}
    \end{subfigure}
    \newline
        \begin{subfigure}[t]{0.45\textwidth}
        \includegraphics[width=\textwidth]{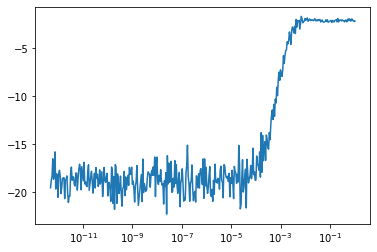}
        \caption{t-statistics.}
    \end{subfigure}
    ~ %add desired spacing between images, e. g. ~, \quad, \qquad, \hfill etc. 
    \begin{subfigure}[t]{0.45\textwidth}
        \includegraphics[width=\textwidth]{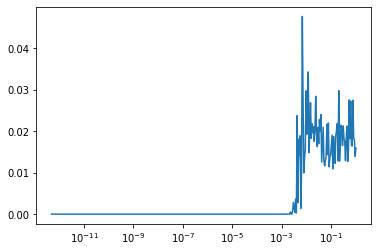}
        \caption{p-values}
    \end{subfigure}
    \caption{The true log ratio $\log Z_{X_{v},\mathcal{M},\mathcal{C},t}$ in orange and classifier-driven approximate in blue for a Gaussian linear regression statistical model with t-noise distributed true linear model, above the t-statistics and corresponding p-values of a one-tailed t-test evaluating whether the sum log ratio is greater than zero.}\label{fig:regressiontnoise}
\end{figure}

Figure \ref{fig:regressiontnoise} shows that CARMEN captures the optimum of the expected log-ratio well, with some bias and noise again becoming evident at larger negative values. Figure \ref{fig:regressiontnoise} also displays the t-statistics and corresponding p-values for different degrees of tempering: the t-statistics appear to plateau for smaller values of tempering as the greater noise in the more negative values of $\log Z_{X_{v},\mathcal{M},\mathcal{C},t}$ counteracts the more extreme mean. For the optimal degree of tempering $t=4.8 \cdot 10^{-3}$ derived from maximising the log generalised posterior predictive, a p-value of 0.018 is returned, indicating borderline evidence of misspecifiation for this model. A borderline p-value is appropriate for an example with a well-specified mean, while it would nevertheless benefit from the heavier tales of a partial update.

\begin{figure}
    \centering
    \begin{subfigure}[t]{0.45\textwidth}
        \includegraphics[width=\textwidth]{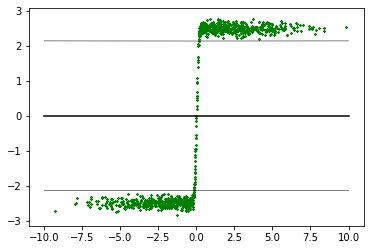}
        \caption{Prior predictive.}
        \label{fig:tiger}
    \end{subfigure}
    ~ %add desired spacing between images, e. g. ~, \quad, \qquad, \hfill etc. 
    \begin{subfigure}[t]{0.45\textwidth}
        \includegraphics[width=\textwidth]{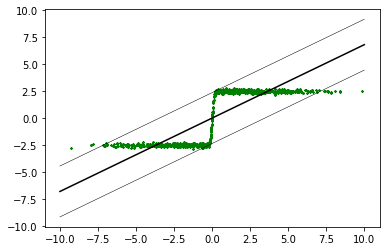}
        \caption{Posterior predictive.}
        \label{fig:mouse}
    \end{subfigure}
    \caption{The predictive prior and predictive posterior distributions of a linear model trained on data with a sigmoid mean function. Means, $5 \% $ and $95 \% $ predictive quantiles are shown as solid lines.}\label{fig:priorpostregressionsigmoid}
\end{figure}

\subsubsection{Linear Regression with sigmoidal truth}\label{sec:regressionsigmoid}

Here we consider the situation in which the true data is generated from a model with sigmoidal mean function and Gaussian distributed noise:

\begin{align}
y_{o}\sim \mathcal{N}(5( \Phi(10X_{o})-.5),.1^2)
\end{align}

where $\Phi$ is the CDF of a standard Gaussian distribution. It is not unrealistic to fit a linear model to a model with such a mean function, as the response variable $y$ is continuously distributed and a positive dependence on the covariate $X$ would be a correct prior assumption from an expert.

The prior model described in Section \ref{sec:regression} was used here and the prior predictive distribution and posterior predictive distributions after a full Bayesian update are displayed in Figure \ref{fig:priorpostregressionsigmoid}. It is clear that the linear model captures the first order correlation structure present in the data, but is limited from capturing the more complex dependence between the variables.

\begin{figure}
    \centering
    \begin{subfigure}[t]{0.45\textwidth}
        \includegraphics[width=\textwidth]{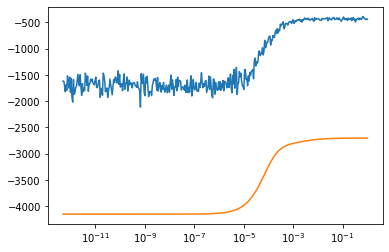}
        \caption{$\log Z_{X_{v},\mathcal{M},\mathcal{C},t}$ for all $t$ up to a full Bayes update.}
    \end{subfigure}
    ~ %add desired spacing between images, e. g. ~, \quad, \qquad, \hfill etc. 
    \begin{subfigure}[t]{0.45\textwidth}
        \includegraphics[width=\textwidth]{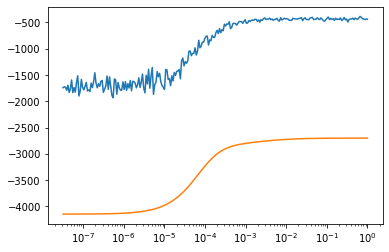}
        \caption{$\log Z_{X_{v},\mathcal{M},\mathcal{C},t}$ for $t$ range around the optimum.}
    \end{subfigure}
    \newline
        \begin{subfigure}[t]{0.45\textwidth}
        \includegraphics[width=\textwidth]{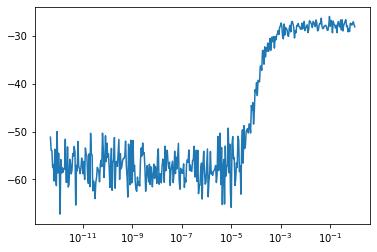}
        \caption{t-statistics.}
    \end{subfigure}
    ~ %add desired spacing between images, e. g. ~, \quad, \qquad, \hfill etc. 
    \begin{subfigure}[t]{0.45\textwidth}
        \includegraphics[width=\textwidth]{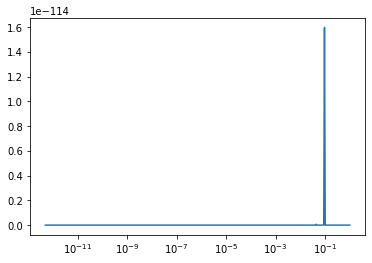}
        \caption{p-values}
    \end{subfigure}
    \caption{The true log ratio $\log Z_{X_{v},\mathcal{M},\mathcal{C},t}$ in orange and classifier-driven approximate in blue for a Gaussian linear regression statistical model with true sigmoid mean model, above the t-statistics and corresponding p-values of a one-tailed t-test evaluating whether the sum log ratio is greater than zero.}\label{fig:regressionsigmoid}
\end{figure}

Logistic regression classifier was used with the summary statistics: $y$, $|y|$, $y^2$, $yX$, $|yX|$ and $(yX)^2$. Results are displayed in Figure \ref{fig:regressionsigmoid}. The analytical log ratio increases smoothly as a function of the tempering, showing no intermediate peak as in previous examples. Based purely on the information from the analytical predictive posterior, the model is indistinguishable from the convergence of a well-specified model.

The classifier approximation to the log ratio captures the structure present in the analytical log ratio, albeit with a clear bias. However, it does correctly identify that the expected log ratio has a large negative value everywhere. The corresponding t-statistics for misspecification all show strong evidence for misspecification for every value of tempering, including the optimal value of $t=1$ corresponding to a full Bayesian update. As such, the bias does not change the conclusion suggested by the diagnostic, as it still assigns a very large negative value of the log ratio under all circumstances. It is likely that the data transformations used are not completely sufficient to describe all of the variation between the simulated and observed data, but they are nevertheless sufficient for predicting the class labels accurately.

\section{Discussion}

In this work, we have demonstrated the use of classifiers to analyse the problem of model misspecification for canonical Bayesian inference problems. As shown, classification methods provide estimates of the KL divergence between the statistical model and the true generative process of the observed data. Furthermore, they can be combined with the SafeBayesian framework to derive optimal generalised Bayesian posteriors by tempering beyond that provided by the likelihood function.

The estimate of the KL divergence remains always conditioned on the discriminative abilities of the classifiers. The choice of summary statistics fed into the linear logistic regression was shown to be important for accurate characterisation of the expected log ratio $\log Z_{X_{v},\mathcal{M},\mathcal{C},t}$. A simple classifier may need some transformation of the data to make the distinct features between the simulated and observed data clear enough. It would be possible to use a more expressive discriminative model to perform ratio estimation, for example a Gaussian Process classifier or a neural network. However, more complex classifiers will cover a larger pool of discriminable models at the expense of needing more training data and losing interpretability.

The use of summary statistics in logistic regression also enables the use of expert knowledge of suspected misspecified features, separately to the expert knowledge encoded in the model or the prior. Including a summary statistic, suspected to encode misspecification information, as a covariate in the classifier is much easier than potentially rebuilding a statistical model integrating the associated expert intuition. It would also be possible to include many data transformations in the logistic classifier and use lasso-type regularisation to determine which are relevant, similar to the approach used in LFIRE \cite{thomas2016likelihood}. In this case, the summary statistics identified as powerful discriminating features could be used to inform design of a new statistical model incorporating properties suggested by the selected summaries.

Even with well-chosen summary statistics, it is still possible for the classifier to become poorly calibrated, especially in situations where the observed and simulated data are easily separated. In these instances the model misspecification should already be clear, while the classification approach should remain well-conditioned for difficult discrimination situations where human intuition may fail. It is to be expected that the KL divergence can be underestimated by the classifier as it is explaining as much variation as possible between the observed and simulated data, while the true model may fall outside the set of models discriminable by the classifier.

In the case of a poorly calibrated classifier as in Section \ref{sec:regressionsigmoid}, the corresponding high variance on the estimate for $\log Z_{X_{v},\mathcal{M},\mathcal{C},t}$ can be integrated further into the probabilistic analysis of whether the model is well-specified. In the current optimistic situation of assuming a well-specified model as a null hypothesis, any noise or underestimation of the expected log ratio will become manifest as a conservative influence in the hypothesis testing framework. We stress that while a small t-test p-value can be interpreted as a sign of misspecifcation, a large p-value is not necessarily confirmation of a well-specified model. A mirror-image testing framework with misspecification assumed as the null hypothesis could be used with inverse estimates to those used in this work, but such a framework might exhibit a bias towards labelling poorly-specified models as well-specified when a weak classifier is used. Philosophically, we advocate starting with optimistic assumptions about model specification that are then put under scrutiny.

Classification is already used in combination with simulated data sets in the context of likelihood-free inference for generative models, within both the statistical and machine learning communities. The inaccessibility of the likelihood function has often been considered a disadvantage, since it is assumed that the likelihood offers all the evidence necessary for a complete statistical analysis. However, in the instance of model misspecification, it is potentially useful to incorporate information beyond the model likelihood when an overly-concentrated posterior is a possibility.

\vspace{5mm}

The authors gratefully acknowledge the support of the European Research Council [742158].

\clearpage

\section{Supplementary Material}
\beginsupplement

\clearpage

\bibliographystyle{plain}
\bibliography{second_writeup}

\end{document}